\documentclass[10pt,letterpaper]{article}
\usepackage[top=0.85in,left=2.75in,footskip=0.75in]{geometry}

\usepackage{amsmath,amssymb}

\usepackage{changepage}

\usepackage{textcomp,marvosym}

\usepackage{cite}

\usepackage{nameref,hyperref}

\usepackage[nopatch=eqnum]{microtype}
\DisableLigatures[f]{encoding = *, family = * }

\usepackage[table]{xcolor}

\usepackage{array}

\newcolumntype{+}{!{\vrule width 2pt}}

\newlength\savedwidth

\raggedright
\usepackage[aboveskip=1pt,labelfont=bf,labelsep=period,justification=raggedright,singlelinecheck=off]{caption}

\makeatletter
\renewcommand{\@biblabel}[1]{\quad#1.}
\makeatother

\usepackage{lastpage,fancyhdr,graphicx}
\usepackage{epstopdf}
\fancyheadoffset[L]{2.25in}
\fancyfootoffset[L]{2.25in}
\begin{document}
\vspace*{0.2in}

\begin{flushleft}
{\Large
\textbf\newline{Graph topological transformations in space-filling\\cell aggregates} 
}
\newline
\\
Tanmoy Sarkar\textsuperscript{1} and Matej Krajnc\textsuperscript{1*}
\\
\bigskip
\textbf{1} Department of Theoretical Physics, Jo\v zef Stefan Institute, Ljubljana, Slovenia
\\
\bigskip

%
%





* matej.krajnc@ijs.si

\end{flushleft}
\section*{Abstract}
Cell rearrangements are fundamental mechanisms driving large-scale deformations of living tissues. In three-dimensional (3D) space-filling cell aggregates, cells rearrange through local topological transitions of the network of cell-cell interfaces, which is most conveniently described by the vertex model. Since these transitions are not yet mathematically properly formulated, the 3D vertex model is generally difficult to implement. The few existing implementations rely on highly customized and complex software-engineering solutions, which cannot be transparently delineated and are thus mostly non-reproducible. To solve this outstanding problem, we propose a reformulation of the vertex model. Our approach, called Graph Vertex Model~(GVM), is based on storing the topology of the cell network into a knowledge graph with a particular data structure that allows performing cell-rearrangement events by simple graph transformations. Importantly, when these same transformations are applied to a two-dimensional~(2D) polygonal cell aggregate, they reduce to a well-known T1 transition, thereby generalizing cell-rearrangements in 2D and 3D space-filling packings. This result suggests that the GVM's graph data structure may be the most natural representation of cell aggregates and tissues. We also develop a \texttt{Python} package that implements GVM, relying on a graph-database-management framework \texttt{Neo4j}. We use this package to characterize an order-disorder transition in 3D cell aggregates, driven by active noise and we find aggregates undergoing efficient ordering close to the transition point. In all, our work showcases knowledge graphs as particularly suitable data models for structured storage, analysis, and manipulation of tissue data.

\section*{Author summary}
Space-filling polygonal and polyhedral packings have been studied as physical models for foams and living tissues for decades. One of the main challenges in the field is to mathematically describe complex topological transformations of the network of cell-cell interfaces that are present during cell rearrangements, accompanying plastic deformations and large-scale cellular flows. Our work addresses this challenge by storing the topology of the network of cell-cell interfaces into a knowledge graph with a specific data structure, uniquely defined by a metagraph. It turns out that this graph technology, also used by tech giants such as Google and Amazon, allows representing topological transformations as graph transformations, that are intuitive, easy to visualize, and straight-forward to implement computationally.


\section*{Introduction}
The mechanical interactions between individual cells and between the cells and their environment play a crucial role in determining the macroscopic properties and behaviors of animal tissues on a large scale~\cite{lecuit07,angelini11,park15,mongera18,etournay15,discher05,fiore20}. During embryonic development, for example, forces generated within cellular cortices drive precise and highly orchestrated active deformations and collective cellular flows~\cite{blankenship06,martin09,rauzi15,streichan18,tomer20,tomer22}. The mechanical forces are transmitted across the tissue through cell-cell contacts, which form a complex spatial network with a dynamically changing topology~\cite{bertet04,rauzi08,rauzi10,lecuit11,lecuit15,curran17,krajnc18,krajnc20,krajnc21,staddon19,cavanaugh22,sknepnek23}. Apart from better understanding biological processes such as development, regeneration and disease~\cite{bertet04, keller05,Ajeti19,janiszewska20}, studying the mechanical properties of cell aggregates also facilitates the development of biomimetic materials and devices for applications in medicine, engineering, and materials science~\cite{discher05, huebsch09}.

Tissue-scale mechanics have been addressed by computational models that represent cells as discrete entities with certain physical properties that phenomenologically describe the mechanics at smaller length scales~\cite{alert20,graner92,glazier93,swat12,bi15,misra16,bi16,barton17,mueller19,kim21}. Recent studies compare various widely used computational models for simulating cell assemblies including cellular automaton, cellular Potts model, phase-field model, particle-based model, deformable-particle model, Voronoi model, and vertex model~\cite{Osborne17, Beatrici23, Keshet20}. While many of these approaches naturally incorporate cell rearrangements--processes that are crucial for capturing a realistic rheological tissue response--handling these processes computationally in the vertex model is quite challenging. In particular, the vertex model describes individual cells as polygons (2D) or polyhedra (3D) and parametrizes their shapes by vertex positions~\cite{honda04,farhadifar07,fletcher14,alt17}. Cell-cell contacts are represented by edges in 2D and polygons in 3D and comprise an intricate network whose connectivity needs to be computationally altered upon every cell-rearrangement event. Current state-of-the-art computational tools for vertex-model simulations offer certain solutions to simplify the implementation of these network-reconnection events, however, these methods are not easily generalizable to 3D~\cite{barton17,sknepnek23}.

Generally, vertex models store the topology and geometry of the tissue in a tabular form within arrays and perform topological transformations of the cell-interface network by updating these arrays according to specified rules~\cite{tyssue,mirams13,sussman17,zhangCode}. Although programming the routines that perform these dynamic array updates is still relatively manageable for planar polygonal cell packings and even for 3D surface packings involving polyhedral prism-like cells~\cite{okuda13,okuda18,sui18,rozman20,rozman21}, developing computer codes for simulations of 3D bulk cell aggregates poses a significantly greater challenge. Indeed, since the pioneering work by Honda {\it et al.}~\cite{honda04}, who first introduced a vertex model of 3D cell aggregates, there have only been a few recent works reporting successful attempts of coding a full 3D vertex model with dynamic cell rearrangements~\cite{okuda22,zhang22,lawsonkeister23}. The difficulty of implementing these rearrangements with the conventional data model raises questions about its suitability and challenges our basic understanding of rearrangements in space-filling packings.

Nevertheless, the vertex model often offers a profound mechanistic understanding of tissue behaviors, which is to a large extent facilitated precisely by the explicit presence of cell boundaries and their ability to remodel. Therefore, despite the challenges associated with cell-boundary tracking during cell rearrangements, the vertex model may still be in many respects advantageous over other models, in which cell boundaries need not be explicitly managed. For instance, the vertex model excels in realistically modeling large-scale three-dimensional tissue deformations and cellular flows, that may be driven by local active forces~\cite{honda04, honda80, honda15, alt17} while, at the same time, capturing certain small-scale structural features, including single-cell shapes and their sidedeness within the aggregate.

To address the challenges associated with the implementation of cell rearrangements in the vertex model, we introduce a reformulation of the vertex model called Graph Vertex Model~(GVM). We discover a particular graph-data model, which allows formulating topological transformations of cell networks as simple graph transformations. The blueprint outlining the relationships among the components is specified through a metagraph which is designed in a manner that topological transformations are themselves represented by graphs. This design not only enhances their intuitive and visual understanding but also simplifies their implementation, making it accessible even to researchers with limited programming expertise. Moreover, we demonstrate that within the GVM's data representation, a T1 transition--the basic rearrangement event in 2D cellular tilings--emerges as a subgraph of graph transformations representing cell rearrangements in 3D packings. This allows developing generalized computational codes that are applicable to both 2D and 3D, suggessting that GVM's data model may be the most natural representation of these systems. As a proof of concept, we develop an open-source \texttt{Python} package \texttt{neoVM}, which implements GVM over a graph database, managed in \texttt{Neo4j}~\cite{neovm,neo4j}. We use our new approach to study order-disorder transition in 3D cell aggregates. We characterize the transition and find aggregates undergoing most efficient ordering in the vicinity of the transition point.

\subsection*{Vertex model}
\begin{figure*}[htb!]
\begin{center}
\includegraphics[]{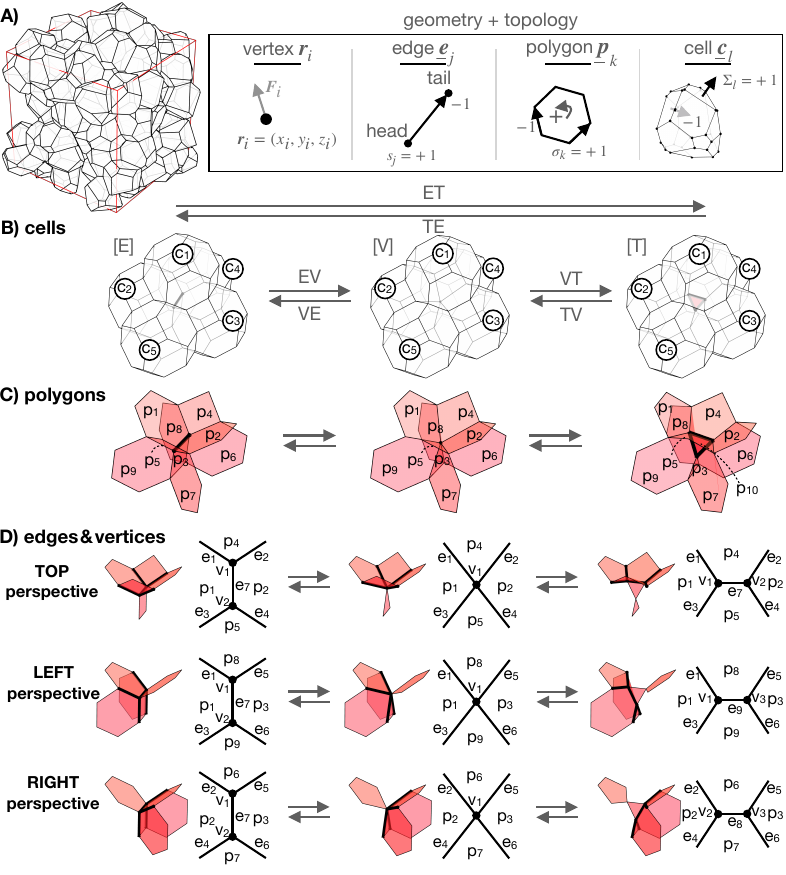}
\caption{\label{F1}\footnotesize\textbf{Geometry and topology of cell aggregates.} {\bf A}~Cell aggregate is modeled by a space-filling packing of polyhedral cells. Cell shapes are parametrized by vertex positions $\boldsymbol r_i$, which move according to mechanical forces $\boldsymbol F_i$. Topology of the cell network is specified by lists $\underline{\boldsymbol e}_j$, $\underline{\boldsymbol p}_k$, and $\underline{\boldsymbol c}_l$ [Eqs.~(\ref{eq:ej})-(\ref{eq:c})], which store head and tail vertices of edges, oriented edges within polygons, and oriented polygons within cells, respectively. {\bf B}~EV transition merges vertices of an edge into a single vertex, whereas VT transition resolves a vertex into a triangle. VE and TV transitions are inverse transitions of EV and VT transitions, respectively. {\bf C}~Polygons involved in topological transitions from panel B. {\bf D}~Decomposed schematic highlighting from 3 different perspectives of polygons, edges, and vertices, in topological transitions from panel B.}
\end{center}
\end{figure*}

The vertex model represents a cell aggregate by a three-dimensional packing of space-filling polyhedral cells~(Fig~\ref{F1}A). Cell shapes are parametrized by positions of vertices in the $(x,y,z)$ space:
\begin{equation}
        \boldsymbol r_i=\left (x_i,\>y_i,\>z_i\right )\>.
\end{equation}
Here $i=1,...,N_v$, where $N_v$ is the total number of vertices.

Pairs of vertices are connected by oriented edges, defined by the indices of the constituent vertices:
\begin{equation}
    \label{eq:ej}
    \underline{\boldsymbol e}_j=\left [+i^{(h)}_j,\>-i^{(t)}_j\right ]\>.
\end{equation}
Here $i^{(h)}_j$ and $i^{(t)}_j$ denote indices of head and tail vertices of edge $j$, respectively~(Fig~\ref{F1}A), and $j=1,...,N_e$, with $N_e$ being the total number of edges. The signs of the vertex indices, $s_j^{(h/t)}=\pm 1$ denote head and tail vertices of the edge. While at this point these signs seem redundant, since the order in which the indices appear in $\underline{\boldsymbol e}_j$ itself indicates the head/tail role of the corresponding vertices, they will become important later on.
    
Polygonal cell sides are defined by oriented lists of indices of their constituent edges as
\begin{equation}
    \label{eq:p}
    \underline{\boldsymbol p}_k=\left [\sigma^{(\mu)}_{k}j^{(\mu)}_{k}\right ]_{\mu=1,...,n_k^{(e)}}\>,
\end{equation}
where $k=1,...,N_p$, with $N_p$ being the total number of polygons. The $n_k^{(e)}$ edges in the list $\underline{\boldsymbol p}_k$ are listed sequentially and $\sigma_k^{(\mu)}=\pm 1$ denotes the orientation of the $\mu$-th edge (with index $j^{(\mu)}_{k}$) within the polygon relative to a chosen positive direction~(Fig~\ref{F1}A).
    
Similarly, cells are defined by oriented lists of indices of the $n_l^{(p)}$ constituent polygons as
\begin{equation}
    \label{eq:c}
    \underline{\boldsymbol c}_l=\left [\Sigma^{(\nu)}_{l}k^{(\nu)}_{l}\right ]_{\nu=1,...,n_l^{(p)}}\>,
\end{equation}    
where $l=1,...,N_c$, with $N_c$ being the total number of cells within the aggregate. Polygon orientations $\Sigma^{(\nu)}_{l}=\pm 1$, where $-1$ and $+1$ correspond to the polygon's normal vector pointing towards the cell center and away from the cell center, respectively~(Fig~\ref{F1}A). The direction of the polygon's normal vector is defined by the right-hand-screw rule, where the fingers curl along the chosen polygon's positive orientation of the bounding edges (Fig~\ref{F1}A). In contrast to the case of polygons, $\underline{\boldsymbol p}_k$, polygon indices in $\underline{\boldsymbol c}_l$ need not be listed in any specific order. 

Exemplary structures of cell aggregates with $N_c=128$ cells are given in the online repository of \texttt{neoVM}~\cite{neovm} and their generation is described in Methods.

The dynamics of cell-shape changes are described by simulating movements of the vertices, driven by mechanical forces. In a model that neglects inertial effects and only considers friction of vertices with a static background, vertices follow first-order dynamics described by 
\begin{equation}
    \label{eq:dynSys}
    \eta\frac{{\rm d}\boldsymbol r_i}{\rm dt}=-\nabla_i W+\boldsymbol F_i^{(a)}\>.
\end{equation}
Here, $\eta$ is the friction coefficient, $W$ is the potential energy of the aggregate, and $\boldsymbol F_i^{(a)}$ is a system-specific active-force contribution. The potential energy is typically calculated from geometric properties of cells such as surface areas of cell-cell contacts $A$, cell volumes $V$, etc. These quantities are calculated from the vertex positions as sums over geometric elements, i.e., vertices, edges, polygons~(Methods).

\subsection*{Topological transitions}
In addition to changing their shapes, cells also change their relative organization within the tissue by exchanging their neighbors~\cite{honda04}. These reorganization events alter the topology of the network of cell-cell contacts, thereby also affecting the interaction between the vertices. In confluent cell aggregates, cells exchange their neighbors by (i)~merging vertex pairs of vanishingly short edges and (ii)~resolving these vertices into new edges~\cite{schwarz64,weaire94,reinelt96}. The topology of the edge network after these transformations locally differs from the initial one. In particular, cells that were initially separated might become neighbors, whereas pairs of initially neigboring cells may separate.

To model cell rearrangements in 3D cell aggregates, the following elementary local topological trasformations need to be considered~(Figs~\ref{F1}B-D): (i)~Edge-to-vertex (EV) transition merges vertices of a vanishingly short edge into a single 6-fold vertex, (ii)~vertex-to-triangle (VT) transition resolves a 6-fold vertex into a new triangle, (iii)~triangle-to-vertex (TV) transition merges all three vertices of a triangular polygon into a single 6-fold vertex, and (iv)~vertex-to-edge (VE) transition resolves a 6-fold vertex into a new edge.

An EV followed by a VT completes an ET transition, which transforms a vanishing edge into a triangle, formed in the perpendicular direction to the shrinking edge~(Fig~\ref{F1}B~and~\ref{F1}C). After an ET transition, initially separated cells become neighbors by sharing the new triangle. Similarly, a TV followed by a VE, completes a TE transition, which transforms a vanishing triangle into an edge, formed in the perpendicular direction to the shrinking triangle~(Fig~\ref{F1}B~and~\ref{F1}C). After a TE transition, the neighboring cells initially sharing the triangle become separated. We note that in a special case of an epithelial monolayer, where cells only attach to their neighbors laterally but not apically and basally, EV transitions either on basal or apical edges give rise to scutoidal cells~\cite{rupprecht17,gomezgalvez18}.

Importantly, as illustrated in Fig~\ref{F1}D, the basic topological transformations in fact resemble multiple edge-to-vertex and vertex-to-edge transitions, known in 2D polygonal networks as T1 transitions. In a T1 transition, a pair of initially neigboring polygons becomes separated, whereas polygons from the remaining (initially separated) polygon pair become neighbors.  This similarity between 2D and 3D transformations suggests that it may be possible to unify topological transformations in 2D and 3D space-filling packings, provided that the vertex model's core architecture be properly reformulated.

The main issue with the conventional implementation of the vertex model is that vertices, edges, polygons, and cells are stored in separate arrays (${\boldsymbol r}_i$, $\underline{\boldsymbol e}_j$, $\underline{\boldsymbol p}_k$, and $\underline{\boldsymbol c}_l$, respectively), which do not directly encode any interconnections or relationships among their respective elements. In particular, elements that might be spatially and topologically related are generally not stored together in the database and accessing any high-level topology data (e.g., finding cells that share a common vertex) requires inefficient searches over all the elements of the cell network. To avoid these inefficient searches, the conventional vertex models store data in a highly redundant form, where higher-level information about the topology of the cell network are stored in addition to $\underline{\boldsymbol e}_j$, $\underline{\boldsymbol p}_k$, and $\underline{\boldsymbol c}_l$ (even though these higher-level information may be calculable from $\underline{\boldsymbol e}_j$, $\underline{\boldsymbol p}_k$, and $\underline{\boldsymbol c}_l$). For instance, to efficiently search for cells that share a certain vertex, lists of cells sharing a common vertex need to be stored for all vertices. Indeed, retreiving this information from $\underline{\boldsymbol e}_j$, $\underline{\boldsymbol p}_k$, and $\underline{\boldsymbol c}_l$ on the fly would require highly inefficient looping over all the cells. Due to this data redundancy, algorithms that manipulate the data arrays upon topological transformations in a self-consistent manner are difficult to program.

\section*{Results}
\subsection*{Knowledge graph}
To overcome challenges related with implementing topological transformations in the vertex model, we propose a new approach, based on storing the topology of the cell network into a knowledge graph, which uses a graph- rather than a tabular data model. By construction, the elements that are topologically related are also connected in the knowledge graph and therefore, any high-level information about the topology of the cell network is readily retrievable by querying over the relevant part of the database with no need of storing any redundant data.

Knowledge graph is a graph data structure, which represents a network of real-world entities and relationships between them~\cite{paulheim17}. These data are stored in a graph database where entities and relationships are represented by nodes and links, respectively, and can, additionally, carry multiple properties. For example, a movie database can be stored as a knowledge graph, in which the data about actors and directors for a given movie are represented by nodes labeled \texttt{Person} and \texttt{Movie} and relationships labeled \texttt{ACTED\_IN} and \texttt{DIRECTED}. In such a knowledge graph, the information that Cillian Murphy acted in the movie Oppenheimer, directed by Christopher Nolan, can be stored as \texttt{(p1:Person \{name: "Cillian Murphy"\})-[:ACTED\_IN]->(m:Movie \{title: "Oppenheimer"\})<-[:DIRECTED]-(p2:Person \{name: "Christopher Nolan"\})}; here nodes labeled \texttt{Person} and \texttt{Movie} carry the \texttt{name} and the \texttt{title} properties, respectively. 

The notation used here and in the following sections follows the syntax of the \texttt{Cypher} graph query language~\cite{cypher}. It is important to note that GVM relies on general principles of discrete mathematics and does not depend neither on the choice of the query language nor on the choice of the database-management framework. Computationally implementing GVM, however, does require some basic knowledge of graph databases and query languages.

\subsection*{Metagraph}
Real-world entities in GVM are vertices, edges, polygons, and cells and they are stored in a knowledge graph in a hierarchical manner as nodes labeled \texttt{Vertex}, \texttt{Edge}, \texttt{Polygon}, and \texttt{Cell}, respectively. The topology of the cell network is encoded through relationships labeled \texttt{IS\_PART\_OF}. These relationships are directed and relate source-target node pairs, where the entity represented by the source node is always hierarchically one level below the entity represented by the target node.

For instance, if a specific polygon $p$ contains a specific edge $e$, the nodes representing these two entities, i.e., \texttt{(p)} and \texttt{(e)}, are connected as \texttt{(e)-[:IS\_PART\_OF]->(p)}. Connecting equally labeled nodes (e.g., a pair of edges) is not allowed and neither is connecting nodes carrying labels that do not follow one another hierarchically (e.g., a vertex-polygon pair). For example, say that one of the polygon $p$'s vertices is vertex $v$. Rather than encoding the information that $v$ is part of $p$ directly by a \texttt{(v)-[:IS\_PART\_OF]->(p)} connection, this information is retrieved hierarchically from the connectivity of $v$ and $p$ through edges. In particular, if $v$ is part of $p$, it is also necessarily shared by two edges that are both also part of $p$ (say $e_1$ and $e_2$): \texttt{(v)-[:IS\_PART\_OF]->(e1)-[:IS\_PART\_OF]->(p)} and \texttt{(v)-[:IS\_PART\_OF]->(e2)-[:IS\_PART\_OF]->(p)}. An extra connection \texttt{(v)-[:IS\_PART\_OF]->(p)} would be redundant and is therefore forbidden in GVM, since these two subgraphs already imply that $v$ is one of polygon $p$'s vertices.

\begin{figure*}[htb!]
\begin{center}
\includegraphics[]{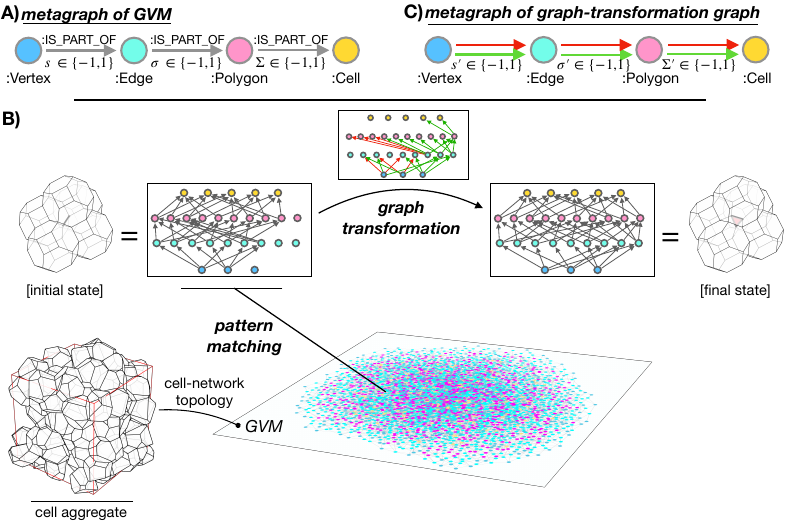}
\caption{\label{F2}\footnotesize\textbf{Graph Vertex Model.} {\bf A}~The hierarchical structure of GVM vertex$\to$edge$\to$polygon$\to$cell is defined by a metagraph. Relationships between these nodes are labeled \texttt{IS\_PART\_OF} and carry a \texttt{sign} property denoted by $s$, $\sigma$, and $\Sigma$ for vertex$\to$edge, edge$\to$polygon, and polygon$\to$cell relationships, respectively. {\bf B}~A subgraph representing a particular local cell state is obtained by pattern matching. This subgraph is then transformed by a graph transformation. {\bf C}~Metagraph of graph-transformation graph connects \texttt{Vertex}, \texttt{Edge}, \texttt{Polygon}, and \texttt{Cell} nodes with green and red relationships, indicating creation and deletion of \texttt{IS\_PART\_OF} relationships in the GVM's knowledge graph, respectively.}
\end{center}
\end{figure*}

The above rules for the construction of the GVM's knowledge graph can be conveniently represented by a graph, called {\it metagraph}. Much like metalanguage is a language that describes another language, metagraph is a graph that describes another graph and can be viewed as a blueprint for generating actual (valid) manifestations of that graph. From the above definitions, it is obvious that the metagraph of GVM is \texttt{(:Vertex)-[:IS\_PART\_OF]->(:Edge)-[:IS\_PART\_OF]->(:Polygon)-[:IS\_PART\_OF]->} \texttt{->(:Cell)}~(Fig~\ref{F2}A).

Additionally, both nodes and relationships carry properties that encode additional information about nodes and relationships. While nodes carry a property, \texttt{id}, which represents the identification numbers $i$, $j$, $k$, and $l$ of vertices, edges, polygons, and cells, respectively, the relationships are prescribed a property \texttt{sign}, whose value is either $-1$ or $+1$. This is a contextual property, that puts the relationship's source node into the context of the target node. In particular, in the subgraphs of type \texttt{(v:Vertex)-[r:IS\_PART\_OF]->(e:Edge)}, the value of \texttt{r.sign} denotes whether vertex $v$ is a head vertex (\texttt{r.sign=+1}) or a tail vertex (\texttt{r.sign=-1}) of edge $e$~[i.e., parameter $s$ in Eq.~(\ref{eq:ej})]. In the subgraphs of types \texttt{(e:Edge)-[r:IS\_PART\_OF]->(p:Polygon)} and \texttt{(p:Polygon)-[r:IS\_PART\_OF]->(c:Cell)} the same property specifies the orientation of edge $e$ in the context of polygon $p$ and polygon $p$ in context of cell $c$, respectively [i.e., parameters $\sigma$ and $\Sigma$ in Eqs.~(\ref{eq:p}) and~(\ref{eq:c}), respectively].

\subsection*{Pattern matching}
Transforming the graph database of GVM upon cell rearrangements requires only two steps: (i) Data retrieval, accomplished through pattern matching and (ii) Graph transformation. 

In step (i), a suitable (meta)graph pattern is utilized to query the database and identify the nodes relevant to the transformation at hand. After the graph is traversed, instances of the specified graph pattern are returned. These instances are further filtered, using various conditional statements.

The goal of step (i) is to retrieve from the whole graph of GVM a small subgraph comprising solely of the nodes representing the objects (vertices, edges, polygons, and cells) that take part in the specific topological transformation being performed~(Fig~\ref{F2}B). Given the unambiguous definition of the graph data structure by the GVM's metagraph~(Fig~\ref{F2}A), the routines that perform this step are easily reproducible. We implement these routines in \texttt{Cypher} and find that each topological transition requires $\sim 10$ distinct short queries, similar to the query shown in Eq.~(\ref{eq:cypher1}) to retrieve the relevant data~\cite{neovm}.

\subsection*{Graph transformations}
In step (ii), the subgraph matched during the pattern-matching step undergoes a transformation based on the rules of the specific cell-rearrangement event being performed. 

Like the matched subgraph that is being transformed, the graph transformation itself is represented by a graph. This graph contains exactly the same nodes as the matched GVM subgraph, however with much fewer relationships. In particular, the relationships in the transformation graph are of two types: (i)~Green and (ii)~red, indicating creations and deletions of \texttt{:IS\_PART\_OF} relationships in the actual GVM subgraph, respectively~(Fig~\ref{F2}C). 

Compared to the convoluted codes that perform topological transformations in the conventional vertex model, the task of programming the routines that perform graph topological transformations in GVM is much less challenging. Indeed, our implementation of graph-transformation routines in \texttt{Cypher} comprises of successive calls of $\sim 5$ distinct short queries, similar to examples shown in Eqs.~(\ref{eq:cypher2}) and (\ref{eq:cypher3}), which merely delete and create relationships.

{\bf Transformation of contextual properties.} Values of contextual properties $s$, $\sigma$, and $\Sigma$ to be assigned to the newly created relationships are specified in the transformation graph through the relationship property of green relationships ($s'$, $\sigma'$, and $\Sigma'$ for vertex$\to$edge, edge$\to$polygon, and polygon$\to$cell connections, respectively). Unlike the green relationships, the red relationships do not carry any additional properties~(Fig~\ref{F2}C).

Transformed values of certain contextual properties can be arbitrarily chosen, however, in most cases they need to be figured out from the known values of other contextual properties. Conveniently, the rules for prescribing these new values can be summarized in 2 compact formulas. In particular, for a new vertex$\to$edge relationship between nodes representing vertex $v_i$ and edge $e_j$, $s'_{i,j}$ is calculated as 
\begin{equation}
    \label{s1}
    s_{i,j}'=s_{k,j}\>,
\end{equation}
where $k$ denotes vertex $v_k$, which was part of edge $e_j$ prior to the transformation (i.e., $s_{k,j}$ denotes contextual property of the relationship between nodes representing vertex $v_k$ and edge $e_j$ prior to the transformation).

For a new edge$\to$polygon relationship between nodes representing edge $e_i$ and polygon $p_j$, $\sigma'_{i,j}$ is calculated as
\begin{equation}
    \label{s2}
    \sigma_{i,j}'=-s_{k,i}'s_{k,l}'\sigma_{l,j}\>.
\end{equation}
Here, vertex $v_k$ is a vertex shared by edges $e_i$ and $e_l$, which are both part of polygon $p_j$. Additionally, among the two vertices of edge $e_i$, vertex $v_k$ is the one that was not part of edge $e_l$ prior to the transformation (i.e., Eq.~(\ref{s2}) contains $s_{k,l}'$ and not $s_{k,l}$). An analogous equation to Eq.~(\ref{s2}) also holds for assigning $\Sigma'_{ij}$ to a newly created polygon $\to$ cell relationship. 

Equations (\ref{s1}) and (\ref{s2}) are demonstrated in the subsequent section for the case of a T1 transformation and their meaning is described in the corresponding figure caption.


\subsection*{T1 graph transformation}
\begin{figure*}[htb!]
\begin{center}
\includegraphics[]{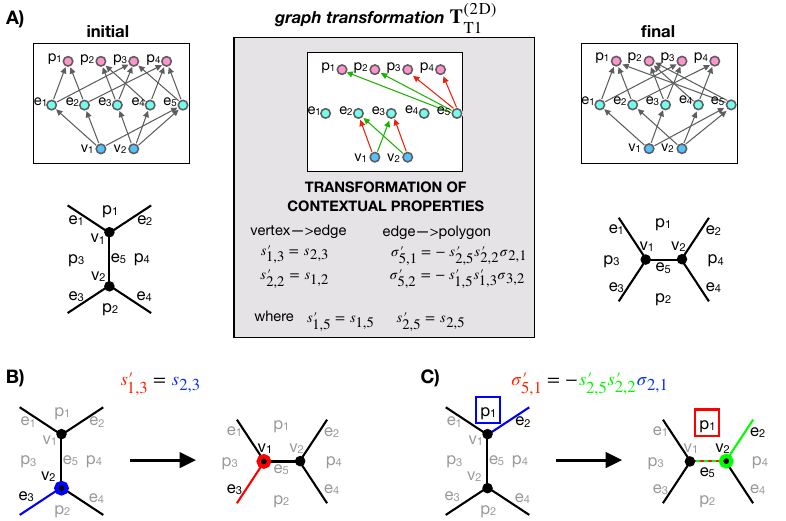}
\caption{\label{F3}\footnotesize\textbf{Graph transformation for a T1 transition.} {\bf A} Graphs in the left and right columns correspond to the initial and final cell configurations, respectively. Gray arrows represent relationships labeled \texttt{IS\_PART\_OF}. The graph in the middle column shows the graph transformation, which includes green and red relationships, indicating relationship creations and deletions, respectively. Additionally, the graph transformation specifies property values of the newly created relationships. {\bf B}~Schematic of determining the value of contextual property for a new vertex$\rightarrow$edge relationship generated between vertex $v_1$ and edge $e_3$ ($s_{1,3}'$). After the transformation, vertex $v_1$ assumes the same role in the context of edge $e_3$ as the role of $v_2$ in the context of $e_3$ before the transformation ($s_{2,3}$). This occurs because the edge $e_3$ merely replaces $v_2$ (blue) with $v_1$ (red). {\bf C} Schematic of determining the value of contextual property for a new edge$\rightarrow$polygon relationship generated between edge $e_5$ and polygon $p_1$ ($\sigma_{5,1}'$). The calculation of $\sigma_{5,1}'$ (red) relies on one of the vertices of $e_5$, i.e., $v_2$ (green), and the edge linked to both $v_2$ and $p_1$, i.e., $e_2$ (also depicted in green). The assignment of $\sigma_{5,1}'$ is determined based on the contextual properties of: (i) edge $e_2$ in the context of polygon $p_1$, $\sigma_{2,1}$, (blue) and (ii) vertex $v_2$ in contexts of edges $e_2$, $s_{2,2}'$, and $e_5$, $s_{2,5}'$, (green). 
In short, the contextual property $\sigma_{5,1}'$ aligns or opposes that of $\sigma_{2,1}$ depending on the similarity or dissimilarity of $s_{2,2}'$ and $s_{2,5}'$.}
\end{center}
\end{figure*}

To demonstrate all steps required to perform a topological transformation in a space-filling cell aggregate, we first turn to a 2D polygonal cellular tiling, where cells rearrange through T1 transitions. Specifically, during a T1 transition, a vanishingly short edge merges into a single vertex, i.e., a four-way junction, which subsequently resolves into a new edge oriented roughly in the perpendicular direction compared to the orientation of the initial edge. As any other topological transformation, GVM performs a T1 transition in two steps as follows~(Fig~\ref{F3}).

(i)~{\bf Pattern matching} performs a series of graph-database queries so as to find nodes representing elements (i.e., vertices, edges, polygons and cells) that participate in the transformation (the initial state in Fig~\ref{F3}A). These queries first identify the edge that undergoes a T1 and label it $e_5$. Subsequently, they identify two vertices connected to $e_5$, randomly labeling one $v_1$ and the other $v_2$. Following this, they locate two polygons that share $e_5$, labeling them $p_3$ and $p_4$. Continuing, edges $e_1$ and $e_2$ are identified as edges connected to $v_1$ (excluding $e_5$) and are also part of polygon nodes $p_3$ and $p_4$, respectively. Similarly, $e_3$ and $e_4$ are determined using the same procedure, where the role of $v_1$ from the previous step is now adapted by $v_2$. Subsequent steps involve finding polygons $p_1$, containing both $e_1$ and $e_2$, followed by identifying $p_2$, containing $e_3$ and $e_4$. 

Note that nodes representing cells (polyhedra) are missing in graphs describing a T1 transition (Fig~\ref{F3}). This is because polyhedra do not exist in the 2D representation of the vertex model, where polygons themselves are interpreted as cells.

(ii) After pattern matching, {\bf graph transformations} convert the initial sub-graph to the final sub-graph, illustrated in the upper-right area of Fig~\ref{F3}A, employing a few transformative operations. The detailed depiction of these graph transformations in the middle panel of Fig~\ref{F3}A reveals several deletions and creations of new relationships. Specifically, at the vertex and edge level,  $v_1 \rightarrow  e_2$ and $v_2 \rightarrow  e_3$ relationships are eliminated, while $v_1 \rightarrow e_3$ and $v_2 \rightarrow e_2$ relationships are established. Furthermore, at edges and polygons level, only $e_5 \rightarrow p_3$ and $e_5 \rightarrow p_4$  relationships are removed, while new $e_5 \rightarrow p_1$ and $e_5 \rightarrow p_2$ relationships are created. What relationships need to be deleted and created is visually represented in the graph-transformations graph by red and green arrows, respectively.

Finally, the newly created relationships need to be prescribed contextual properties using equations (\ref{s1}) and (\ref{s2}),~(Fig~\ref{F3}B~and~\ref{F3}C). For instance, $s_{1,3}'$, i.e., the sign of vertex $v_1$ in the context of the edge $e_3$, is assigned a value identical to $s_{2,3}$, i.e., the sign of vertex $v_2$ in the context of $e_3$ before transformation. In turn, determining the sign of a new edge in the context of a polygon, e.g., $e_5$ in the context of $p_1$ ($\sigma_{5,1}'$), relies on contextual properties $s_{2,5}'$, $s_{2,2}'$, and $\sigma_{2,1}$~[Eq.~(\ref{s2})].
For a more comprehensive understanding of the origin of equations (\ref{s1}) and (\ref{s2}), refer to Fig~\ref{F3}'s caption.

Note that all the graphs shown in Fig~\ref{F3} are unique in a sense that their connectivity does not depend on the labeling of vertices, edges, and polygons. Of course, relabeling these elements or repositioning the corresponding nodes would affect the visual representation of the graphs, but the graphs themselves (i.e., their connectivities) would remain the same.

\subsection*{ET and TE graph transformations}
\begin{figure*}[htb!]
\begin{center}
\includegraphics[]{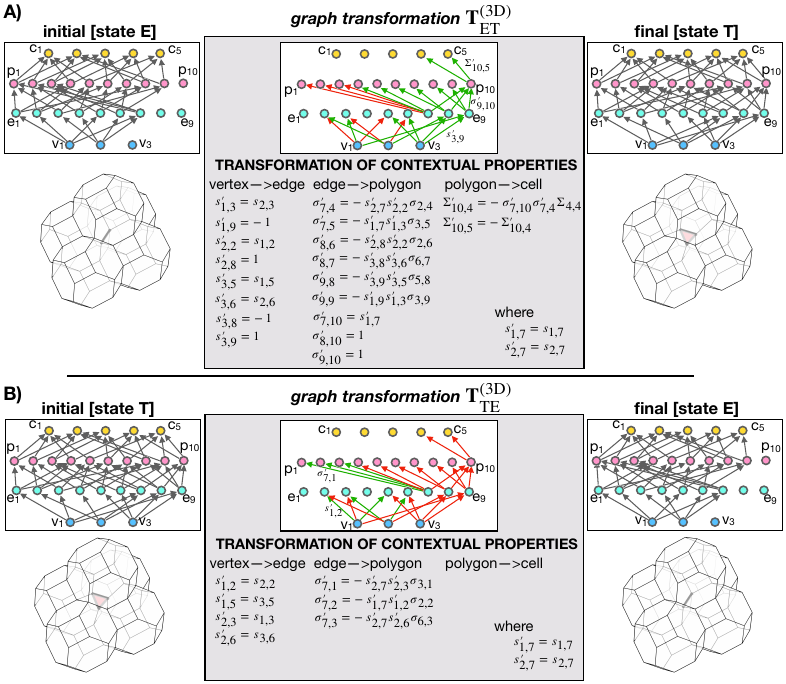}
\caption{\label{F4}\footnotesize\textbf{Graph transformations in a 3D vertex model of polyhedral packings.} {\bf A}~Graph transformation of an ET transition. {\bf B}~Graph transformation of an TE transition. In both panels, graphs in the left and the right column correspond to the initial and the final cell configuration, respectively. Gray arrows represent relationships labeled \texttt{IS\_PART\_OF}. The graphs in the middle column show graph transformations, which include green and red relationships, indicating relationship creations and deletions, respectively. Additionally, the graph transformation specifies property values of the newly created relationships. In each graph, the node indices increase from left to right in unit steps.}
\end{center}
\end{figure*}

Graph transformations describing ET and TE transitions as well as EV, VT, TV, and VE transitions are obtained following the exact same procedure as in the case of T1 transition (previous section). Fig~\ref{F4} shows graph transformations for ET and TE transformations as well as the matched subgraphs representing initial and final cell configurations. Graph transformations for EV, VT, TV, and VE transformations are given in S1 and S2~Figs. For clarity, the graphs representing all three states (E, T, and V) are additionally specified in a more explicit (non-pictorial) form in Methods.

\newpage
\subsection*{Generalization of topological transformations}
As shown in Fig~\ref{F1}D, the transformation patterns during ET and TE transformations are both geometrically as well as topologically quite similar to the more simple T1 transition. This raises a question whether topological transformations in 2D and 3D can be generalized and described using the same, generalized, graph transformation.

\begin{figure*}[htb!]
\begin{center}
\includegraphics[]{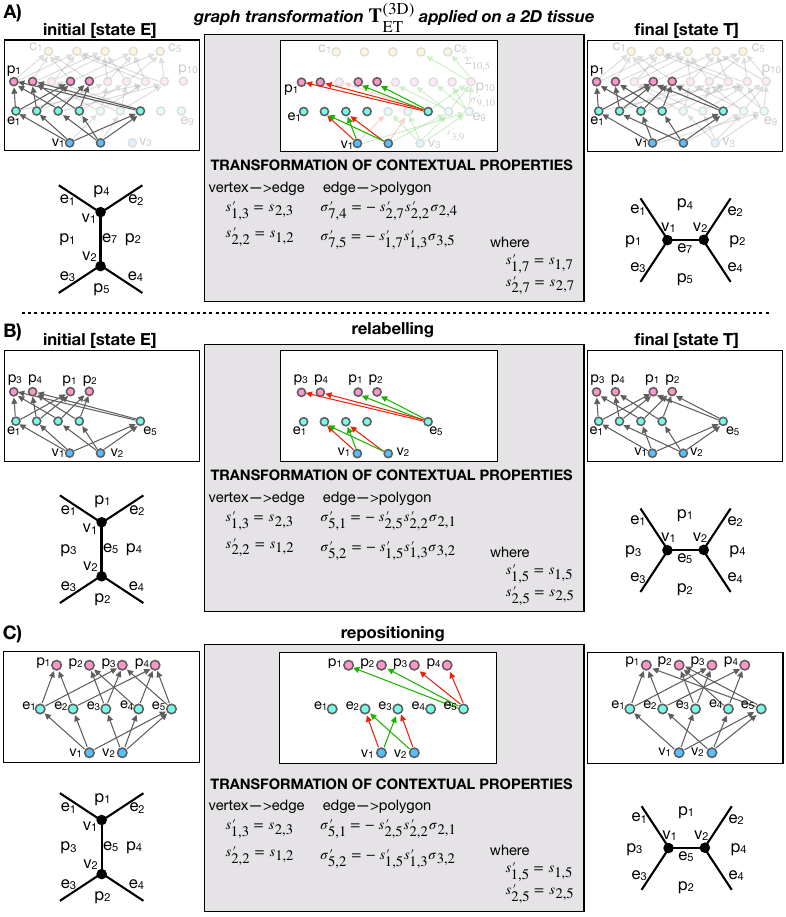}
\caption{\label{F5}\footnotesize\textbf{3D graph transformations reduce to a T1 transition when applied to a 2D vertex model.} {\bf A}~ET transformation when applied to a four-polygon neighborhood described by a 2D vertex model. Parts of graphs shown in transparent are not matched because the corresponding elements do not exist in 2D. After relabeling the nodes (panel {\bf B}) and repositioning them (panel {\bf C}), we observe that the matched graph transformation exactly corresponds to ${\bf T}_{\rm T1}^{\rm (2D)}$, i.e., the graph transformation that performs a T1 transition~(Fig~\ref{F3}).}
\end{center}
\end{figure*}

Surprisingly, as depicted in Fig~\ref{F5}, our Graph vertex model readily resolves this question. Indeed, when either ET or TE transformation (Fig~\ref{F5} only shows the case of ET transformation) are applied on a 2D GVM of polygonal cell aggregates, only a part of the initial subgraph is matched, whereas the rest (shown in transparent in Fig~\ref{F5}) corresponds to elements (vertices, edges, polygons, and cells) that do not exist in the 2D model due to the reduced dimensionality. In turn, only the matched subgraph gets transformed and by relabelling and repositioning the nodes, we prove that the remaining graph transformation exactly corresponds to the graph transformation ${\bf T}_{\rm T1}^{\rm (2D)}$~(Fig~\ref{F3}A), describing a T1 transition. In short, T1 graph transformation can be viewed as a subgraph of both ET and TE graph transformations. 

This result generalizes topological transformations in 2D and 3D vertex models, suggesting that it should be possible to develop a generalized computational implementation of GVM, capable of simulating both 2D and 3D space-filling packings. Our implementation of GVM within the \texttt{neoVM} package confirms this hypothesis.

\subsection*{Order-disorder transition in active tissues}
As a proof of concept, we develop a custom \texttt{Python} package called \texttt{neoVM}, which manages the GVM's knowledge graph and its transformations in a graph database management framework \texttt{Neo4j}~(Methods). 

We use \texttt{neoVM}, to study an order-disorder transition of cell aggregates, driven by active tension fluctuations. In particular, we consider an aggregate of $N_c=128$ cells with identical (normalized) volumes ($V_0=1$ for all cells), enclosed within a simulation box with periodic boundary conditions. The vertex dynamics are described by Eq.~(\ref{eq:dynSys}), assuming the potential energy given by Eq.~(\ref{energyEq1}).

In addition to the conservative and friction forces, we also include active force dipoles acting along cell edges to induce active cell rearrangements. The total active force on vertex $i$ is a sum of forces acting along edges (i.e., tricellular junctions) sharing that same vertex:
\begin{equation}
    \boldsymbol F_i^{(a)}=-\sum_{lmn}\gamma_{lmn}(t)\nabla_i L_{lmn}\>.
\end{equation}
Here the indices $lmn$ denote a tricellular junction (i.e., edge), shared by cells $l$, $m$, and $n$; $L_{lmn}$ is the edge length.

The magnitudes of active force dipoles are dynamic quantities that fluctuate with time. In particular $\gamma_{lmn}(t)$ obeys Ornstein-Uhlenbeck dynamics described by
\begin{equation}
    \label{eq:ou}
    \dot\gamma_{lmn}(t)=-\frac{1}{\tau_m}\left (\gamma_{lmn}(t)-\gamma_0\right )+\xi_{lmn}(t)\>,
\end{equation}
where $\tau_m$ is the relaxation time scale associated with turnover dynamics of molecular motor Myosin, $\gamma_0$ is a baseline tension, whereas $\xi_{lmn}(t)$ is Gaussian white noise with properties $\langle\xi_{lmn}(t)\rangle=0$ and $\langle\xi_{lmn}(t)\xi_{opr}(t')\rangle=\left (2\sigma^2/\tau_m\right )\delta_{lo}\delta_{mp}\delta_{nr}\delta(t-t')$; $\sigma^2$ is the long-time variance of the tension fluctuations.

\begin{figure*}[htb!]
\includegraphics[]{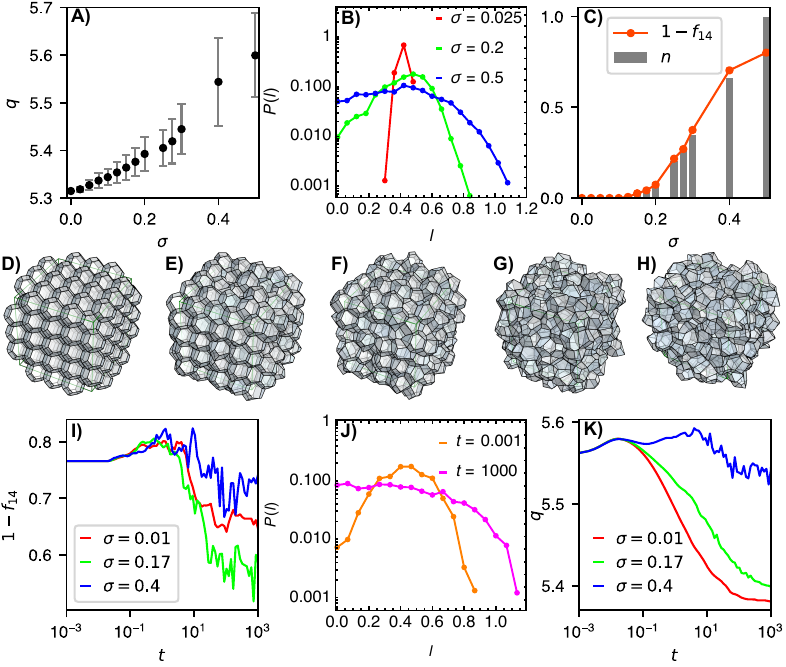}
\caption{\label{F6}\footnotesize\textbf{Order-disorder transition in 3D cell aggregates.} {\bf A}~Shape parameter $q$ for ``thermalized'' aggregates versus active noise $\sigma$. {\bf B}~Distribution of edge lengths in thermalized aggregates for $\sigma=0.025,\>0.2,\>{\rm and \hspace{0.1cm}}\>0.5$ (red, green, and blue curves, respectively). {\bf C}~Order parameter $1-f_{14}$ and normalized number of cell-rearrangement events $n=(N_{\rm ET}+N_{\rm TE})/N_{\rm max}$ versus active noise $\sigma$. {\bf D-H}~Thermalized aggreages at $\sigma=0, 0.2, 0.3, 0.4, {\rm and \hspace{0.1cm}} 0.5$. {\bf I}~Order parameter $1-f_{14}$ versus time for $\sigma=0.01,\>0.17\>{\rm and}\>0.4$ (red, green, and blue curves, respectively) during aggregate ordering. {\bf J}~The initial ($t=0.001$; orange curve) and final ($t=1000$; magenta curve) distributions of edge lengths at $\sigma=0.17$. {\bf K}~Shape parameter $q$ versus time for $\sigma=0.01,\>0.17\>{\rm and}\>0.4$ (red, green, and blue curves, respectively) during aggregate ordering.}
\end{figure*}

We simulate the above active dynamics at different magnitudes of active noise $\sigma$, starting with a Kelvin structure--a crystalline cell arrangement made up of truncated octahedra with 14 facets (8 regular hexagons and 6 squares). The active noise distorts the geometry of the aggregates, which are no longer perfect crystals. In particular, for $\sigma>0$ the average cell shape, quantified by the shape factor $q=\langle S_l/V_l^{2/3}\rangle_{l\in{\rm cells}}$ ($S_l$ and $V_l$ are cell surface area and volume, respectively) deviates from the Kelvin's truncated octahedron~(Fig~\ref{F6}A). The distorted geometry is also seen in the width of the distribution of edge lengths, which increases with an increasing $\sigma$--to a point where vanishingly short edges appear~(Fig~\ref{F6}B). These edges undergo ET and TE topological transformations, which in turn triggers cell rearrangements--a signature of a transition from a solid-like to a fluid-like behavior. This transition manifests in disordering of the aggregate, seen in the dependence of an order parameter $1-f_{14}$, describing the fraction of non-14-sided polyhedra ($f_{14}=N_{14}/N_c$), on the control parameter $\sigma$~(Fig~\ref{F6}C). From these results, we obtain an estimate for the transition point $\sigma^*\approx 0.17$. Figs~\ref{F6}D-H show cell configurations at $\sigma=0, 0.2, 0.3, 0.4, {\rm and \hspace{0.1cm}} 0.5$.

Next, we are interested in whether the active noise can drive the opposite effect, i.e., ordering. To study this, we start with a disordered cell packing, prepared in advance by packing spheres in the simulation box, using random sequential addition and then constructing Voronoi partitions around sphere centers. This procedure yields a sample with the initial fraction of 14-sided polyhedra $f_{14}=0.234$. Again, we simulate the active dynamics at different magnitudes of the active noise $\sigma$. We find that high-$\sigma$ values keep the aggregate disordered due to frequent cell-neighbor exchanges. In contrast, a sufficiently small level of active noise drives active tissue ordering, which is seen in decreasing $1-f_{14}$ and narrowing of the edge-length distribution over time~(Fig~\ref{F6}I and~\ref{F6}J, respectively). At moderate $\sigma\approx\sigma^*$ values, the ordering is more efficient compared to small $\sigma$ values, where the active-noise level is not sufficiently high to allow overcoming local energy barriers for cell rearrangements. Despite a higher degree of disorder, the low-$\sigma$ states consist of cells whose shapes are closer to the Kelvin's regular truncated octahedron compared to cells from the $\sigma\approx\sigma^*$ case, where cell shapes are perturbed due to active noise~(Fig~\ref{F6}K).

The rate of topological-transition events in the simulations of active cell aggregates reaches as high as $\sim 200/{\rm time\>unit}$, which in total amounts to $\sim10^5$ events per simulation~(Fig~\ref{F6}C); note that many of these events are reversible transitions that occur multiple times while the manipulated vertices are still located very close to one another and have not yet properly resolved in the geometric sense. Despite this large number of reconnections, the aggregate does not develop any nonphysical topology, e.g., an edge connecting more than two vertices or a polygon with one missing edge, etc. This demonstrates that topological transformations in space-filling 3D packings can indeed be implemented as graph transformations defined in Fig~\ref{F4}. Importantly, due to GVM's unambiguous data structure, these transformations are relatively straight-forward to implement and are therefore readily reproducible, which is clearly demonstrated by the implementation of GVM within our \texttt{neoVM} package~\cite{neovm}.

\section*{Discussion}
We reformulated the vertex model of cell aggregates. The new formulation, called Graph Vertex Model~(GVM), is based on storing the topology of the cell network into a knowledge graph. We discovered a particular graph data model, uniquely defined by a metagraph, which allows formulating topological transformations of the cell network as simple and mathematically properly defined graph transformations~(Fig~\ref{F2}). These transformations are themselves represented by graphs and consist of only the most elementary graph operations, i.e., relationship deletions and creations.

We designed graph transformations for all topological transitions which are required to describe cell rearrangements--including edge-to-triangle (ET) and triangle-to-edge (TE) transformations~(Fig~4). Importantly, ET and TE transformations can be both applied to a 2D system, where they reduce to the well-known T1 transition. We showed that this happens because the transformation graph describing a T1 transition is in fact a subgraph of both ET and TE transformation graphs~(Figs~\ref{F5} and \ref{F3}). Thus, when ET or TE transformations are applied onto a 2D polygonal tiling, only the T1 subgraph is matched and the corresponding transformation executed. This result generalizes topological transformations in 2D and 3D space-filling packings, suggesting that our proposed graph-data structure may be the most natural representation of the topology of space-filling packings. We used GVM to study active cell aggregates, whose cell junctions are subject to fluctuating active tensions~(Fig~\ref{F6}). In particular, we characterized the order-disorder transition and found initially disordered aggregates undergoing ordering, which is most efficient for active noise close to the transition point.

Even though the basic GVM's data model presented here only encodes information on the topology of the network of cell-cell interfaces, GVM already represents an important technological and conceptual step forward in computational models of tissue mechanics. This advancement lays the foundation for the development of knowledge graphs capable of structurally storing live-imaging data, such as that obtained from developing embryos, integrating data on geometry, topology, mechanics, and biochemistry. With this aim, our ongoing work uses GVM as a starting point to develop a comprehensive knowledge-graph database of the early fly development. Making this database interactive and accessible online will allow collaborative research with the aim to progressively expand our collective knowledge base about the mechanics of the embryonic development. Additionally, its graph data structure may even be readily complemented with graph-compatible methods of artificial intelligence (e.g., Graph Neural Networks).

Graph vertex model can be readily extended to describe other cell-scale events that change the local topology of the network of cell-cell interfaces. Indeed, integrating cell division and apoptosis into the model would allow detailed mechanistic studies of spheroid-like cell aggregates as models of tumors or even embryos during the early stages of development, where cells may still be packed within a three-dimensional aggregate~\cite{liu2022,ichbiah23}. In the context of tumors, for instance, investigating how the stability of the overall tumor shape depends on smaller-scale biomechanical processes such as the cells' effective surface tension, inhomogeneous cell proliferation, and active noise at cell-cell interfaces, would allow better understanding the mechanical basis of tumors' transition to malignancy.

For more efficient simulations, capable of dealing with hundreds or even thousands of cells, a considerable effort will also have to be devoted to developing technologies that will improve computational efficiency of the vertex-model simulations over a graph database. While our current implementation of GVM, \texttt{neoVM}~\cite{neovm}, primarily serves as a proof of concept, it falls short on the efficiency. The reason for this is mostly twofold: (i)~The time integration of the dynamical system is performed in \texttt{Python}, which is generally slower than some low-level programming languages, and (ii)~Performing operations on the graph database managed by \texttt{Neo4j} necessitates reading from and writing to the local hard drive during cell rearrangements, where the database is stored. That this indeed limits the performance of \texttt{neoVM}, is also seen in S4~Fig, which shows the simulation time depending on the system size. While the dependence itself is linear, tissue activity shifts this linear dependence towards a higher offset value following a larger number of cell-rearrangement events. This shows that cell rearrangements consume most of the simulation time of \texttt{neoVM}, suggesting that the efficiency of \texttt{neoVM} could be significanlty improved by relying on memory storage instead of accessing the local hard drive.

\section*{Materials and methods}
{{\bf Initial configurations.} Initial configurations (.vt3d files) used in simulations (Figs~6 and S4) are available in the git repository of neoVM. Among these initial files, the Kelvin initial configuration is generated by taking a unit cell from Surface evolver~\cite{brakke92} and replicating them in all directions until the number of cells in the aggregate reaches 128. On the other hand, the disordered initial conditions are generated by putting spheres in the simulation box using random sequential addition and, subsequently, constructing Voronoi cells around these spheres using the Voro++ package~\cite{rycrof09}.}

{\bf Calculation of forces.} We neglect the inertial effects so that the friction force $\boldsymbol F_i^{(f)}$ needs to counterbalance the sum of conservative and active forces, $\boldsymbol F_i^{(c)}$ and $\boldsymbol F_i^{(a)}$, respectively. This implies $\boldsymbol F_i^{(f)}+\boldsymbol F_i^{(c)}+\boldsymbol F_i^{(a)}=0$, where the conservative force $\boldsymbol F_i^{(c)}=-\nabla_i W$, with $W$ being the potential energy of the system, whereas $\boldsymbol F_i^{(f)}=-\eta\dot{\boldsymbol r}_i$. Here, only friction with a static background is considered, $\eta$ being the associated friction coefficient. The active force $\boldsymbol F_i^{(a)}$ can describe different system-specific active mechanisms, e.g., active contractions of the cell membrane due to the activity of the underlying cell cortex or traction forces~\cite{curran17,krajnc20,trepat09,bi16}. This model yields a system of first-order dynamic equations for vertex positions given by Eq.~(5).

We consider a model, in which cell-cell interfaces are prescribed by effective surface tensions $\Gamma_{lm}$, which include contributions of the cell cortical tension and cell-cell adhesion~\cite{derganc09,manning10,hannezo13}; the notation $lm$ denotes index of a polygon shared by cells $l$ and $m$. In this model, the total potential energy of the cell aggregate reads
\begin{equation}
    \label{energyEq1}
	W=\sum_{\left<lm\right >}\Gamma_{ lm}A_{lm}+\kappa_V\sum_{l}\left (V_l-V_0\right )^2\>, 
\end{equation}
where the first sum goes over all pairs of neighboring cells $l$ and $m$ and the second sum goes over all the cells, $\kappa_V$ and $V_0$ being the cell-incompressibility constant and the preferred cell volume, respectively. Note that $V_0$ need not be equal for all cells. In heterogeneous cell aggregates such as tumors, a similar same energy functional could be used, but $V_0$'s would need to be considered distributed.

By definition, the conservative force acting on vertex $i$ is calculated as
\begin{equation}
	\boldsymbol F_i^{(c)}=-\sum_{\left <lm\right >}\Gamma_{lm}\nabla_i A_{lm}-2\kappa_V\sum_{l}(V_l-V_0)\nabla_iV_l\>,
\end{equation}
which further requires calculating gradients of interfacial surface areas and cell volumes as described below.

Surface area of polygonal side $k$ is calculated as a sum of surface areas of triangular surface elements, $\big\lVert \boldsymbol a_{lm,\mu} \big\rVert$, defined by pairs of consecutive polygon vertices $\boldsymbol r_\mu$ and $\boldsymbol r_{\mu+1}$, and the polygon's center of mass 
\begin{equation}
    \boldsymbol c_{lm}=\frac{1}{n_{lm}}\sum_{\mu\in \langle lm\rangle }\boldsymbol r_\mu\>.
\end{equation} 
Like in the previous section, Greek indices here do not denote the real vertex identification numbers, but their sequential indices within individual polygons. The surface area of polygon $lm$ reads
\begin{equation}
	A_{lm}=
 \sum_{\mu\in \langle lm\rangle} \frac{1}{2}\big\lVert \left (\boldsymbol r_\mu-\boldsymbol c_{lm}\right )\times\left (\boldsymbol r_{\mu+1}-\boldsymbol c_{lm}\right ) \big\rVert
\end{equation}
and its gradient
\begin{align}
    \nabla_i A_{lm}&=\sum_{\mu\in \langle lm\rangle} \frac{1}{2}\nabla_i\Big\lVert\left (\boldsymbol r_\mu-\boldsymbol c_{lm}\right )\times\left (\boldsymbol r_{\mu+1}-\boldsymbol c_{lm}\right )\Big\rVert=\\
    &=\sum_{\mu\in lm}\frac{\big[(\delta_{i\mu} - n_{lm}^{-1})(\mathbf{r}_{\mu+1} - \mathbf{c}_{lm}) + (n_{lm}^{-1}- \delta_{i(\mu+1)})(\mathbf{r}_\mu - \mathbf{c}_{lm}) \big] \times \boldsymbol a_{lm,\mu}}{4\big\lVert \boldsymbol a_{lm,\mu} \big\rVert}\>,
\end{align}
where $\delta_{ij}$ is the Kronecker delta.

Cell volume $l$ is calculated as a sum of volumes of tetrahedra, defined by triangular surface elements $(\boldsymbol r_\mu,\boldsymbol r_{\mu+1},\boldsymbol c_{lm})$ and with the fourth vertex at the origin $(0,0,0)$, as
\begin{equation}
	V_l=\sum_m\sum_{\mu\in \langle lm\rangle}\frac{1}{6}\boldsymbol c_{lm}\cdot\left (\boldsymbol r_\mu\times\boldsymbol r_{\mu+1}\right )\>.
\end{equation}
Its gradient is calculated as
%

\begin{align}
       \nabla_i V_l&=\sum_m\sum_{\mu\in \langle lm\rangle}\frac{1}{6}\nabla_i\left [\boldsymbol c_{lm}\cdot\left (\boldsymbol r_\mu\times\boldsymbol r_{\mu+1}\right )\right ] \nonumber\\
        &=\sum_m\sum_{\mu\in \langle lm\rangle}\frac{1}{6 n_{lm}} (\mathbf{r}_\mu \times \mathbf{r}_{\mu+1}) + \frac{1}{6} (\delta_{i\mu} \mathbf{r}_{\mu+1} - \delta_{i(\mu+1)} \mathbf{r}_\mu) \times \mathbf{c}_{lm}\>.
\end{align}

{\bf Topological transformations.} None of the topological transitions is allowed if the resulting cell configuration breaks any of the following topological rules~\cite{okuda13,zhang22}: (i)~Edge pairs may not share more than one vertex, (ii)~polygon pairs may not share more than one edge and (iii)~cell pairs may not share more than one polygon. In our implementation of GVM within the \texttt{neoVM} packing, these conditions seem to be rigorously imposed, since we never observe them being violated despite conducting numerous simulations involving a large number of rearrangement events~(Fig~6). If such violation were to happen, the transformation causing it would need to be immediately reversed.

{\bf Cypher queries for pattern matching and graph transformations.} The following Cypher query retrieves nodes representing the polygons that share common edge $i$
\begin{equation}
\label{eq:cypher1}
\begin{aligned} 
&\texttt{MATCH (e:Edge)} \\
&\texttt{WHERE e.id=i} \\
&\texttt{MATCH (e)-[:IS\_PART\_OF]->(p:Polygon)} \\
&\texttt{RETURN p}
\end{aligned}
\end{equation}

The following Cypher query creates a new relationship \texttt{IS\_PART\_OF} between nodes \texttt{(v1)} and~\texttt{(e3)} and assigns property value \texttt{s23}.
\begin{equation}
\label{eq:cypher2}
\begin{aligned} 
& \texttt{CREATE (v1)-[:IS\_PART\_OF \{sign:\$s23\}]->(e3)}
\end{aligned}
\end{equation}

The following Cypher query deletes relationship \texttt{r} between nodes \texttt{(e5)} and~\texttt{(p4)}.
\begin{equation}
\label{eq:cypher3}
\begin{aligned} 
& \texttt{MATCH (e5)-[r:IS\_PART\_OF]->(p4)} \\
& \texttt{DELETE r }
\end{aligned}
\end{equation}

{\bf List of relationships.} For clarity, we here explicitly list relationships in graphs representing states E, T, and V~(Figs~4, 5, S1 and S2). All relationships are of type \texttt{IS\_PART\_OF}.
\begin{itemize}
\item {\bf [state E]}: $v_1\to e_1$, $v_1\to e_2$, $v_1\to e_5$, $v_1\to e_7$, $v_2\to e_3$, $v_2\to e_4$, $v_2\to e_6$, $v_2\to e_7$, $e_1\to p_1$, $e_1\to p_4$, $e_1\to p_8$, $e_2\to p_2$, $e_2\to p_4$, $e_2\to p_6$, $e_3\to p_1$, $e_3\to p_5$, $e_3\to p_9$, $e_4\to p_2$, $e_4\to p_5$, $e_4\to p_7$, $e_5\to p_3$, $e_5\to p_6$, $e_5\to p_8$, $e_6\to p_3$, $e_6\to p_7$, $e_6\to p_9$, $e_7\to p_1$, $e_7\to p_2$, $e_7\to p_3$, $p_1\to c_1$, $p_1\to c_2$, $p_2\to c_1$, $p_2\to c_3$, $p_3\to c_2$, $p_3\to c_3$, $p_4\to c_1$, $p_4\to c_4$, $p_5\to c_1$, $p_5\to c_5$, $p_6\to c_3$, $p_6\to c_4$, $p_7\to c_3$, $p_7\to c_5$, $p_8\to c_2$, $p_8\to c_4$, $p_9\to c_2$, $p_9\to c_5$
\item {\bf [state T]}: $v_1\to e_1$, $v_1\to e_3$, $v_1\to e_7$, $v_1\to e_9$, $v_2\to e_2$, $v_2\to e_4$, $v_2\to e_7$, $v_2\to e_8$, $v_3\to e_5$, $v_3\to e_6$, $v_3\to e_8$, $v_3\to e_9$, $e_1\to p_1$, $e_1\to p_4$, $e_1\to p_8$, $e_2\to p_2$, $e_2\to p_4$, $e_2\to p_6$, $e_3\to p_1$, $e_3\to p_5$, $e_3\to p_9$, $e_4\to p_2$, $e_4\to p_5$, $e_4\to p_7$, $e_5\to p_3$, $e_5\to p_6$, $e_5\to p_8$, $e_6\to p_3$, $e_6\to p_7$, $e_6\to p_9$, $e_7\to p_4$, $e_7\to p_5$, $e_7\to p_{10}$, $e_8\to p_6$, $e_8\to p_7$, $e_8\to p_{10}$, $e_9\to p_8$, $e_9\to p_9$, $e_9\to p_{10}$, $p_1\to c_1$, $p_1\to c_2$, $p_2\to c_1$, $p_2\to c_3$, $p_3\to c_2$, $p_3\to c_3$, $p_4\to c_1$, $p_4\to c_4$, $p_5\to c_1$, $p_5\to c_5$, $p_6\to c_3$, $p_6\to c_4$, $p_7\to c_3$, $p_7\to c_5$, $p_8\to c_2$, $p_8\to c_4$, $p_9\to c_2$, $p_9\to c_5$, $p_{10}\to c_4$, $p_{10}\to c_5$
\item {\bf [state V]}: $v_1\to e_1$, $v_1\to e_2$, $v_1\to e_3$, $v_1\to e_4$, $v_1\to e_5$, $v_1\to e_6$, $e_1\to p_1$, $e_1\to p_4$, $e_1\to p_8$, $e_2\to p_2$, $e_2\to p_4$, $e_2\to p_6$, $e_3\to p_1$, $e_3\to p_5$, $e_3\to p_9$, $e_4\to p_2$, $e_4\to p_5$, $e_4\to p_7$, $e_5\to p_3$, $e_5\to p_6$, $e_5\to p_8$, $e_6\to p_3$, $e_6\to p_7$, $e_6\to p_9$, $p_1\to c_1$, $p_1\to c_2$, $p_2\to c_1$, $p_2\to c_3$, $p_3\to c_2$, $p_3\to c_3$, $p_4\to c_1$, $p_4\to c_4$, $p_5\to c_1$, $p_5\to c_5$, $p_6\to c_3$, $p_6\to c_4$, $p_7\to c_3$, $p_7\to c_5$, $p_8\to c_2$, $p_8\to c_4$, $p_9\to c_2$, $p_9\to c_5$
\end{itemize}

{\bf Implementation in Python and Neo4j.} As a proof of concept, we set up the GVM's knowledge-graph database in a graph database management framework \texttt{Neo4j}~\cite{neo4j}. The core program of the vertex model is implemented in \texttt{Python} and communicates with \texttt{Neo4j} using \texttt{Py2neo} client library. The time integration of the dynamical system is performed in \texttt{Python}, whereas all topological transformations are performed in \texttt{Neo4j} through pattern matching and graph transformations implemented as \texttt{Cypher} queries~\cite{cypher}. Our implementation of GVM is available as an open-source \texttt{Python} package, called \texttt{neoVM}, and is available online~\cite{neovm}.

S3~Fig shows the schematic of \texttt{neoVM}'s architecture. The program is initialized by reading the initial geometry and topology of the cell network from an input \texttt{.vt3d} file and storing them into an object \texttt{t} of class \texttt{tissue}. In particular, this object stores lists of \texttt{vertex}, \texttt{edge}, \texttt{polygon}, and \texttt{cell} objects, which encode $\underline{\boldsymbol r}_i$, $\underline{\boldsymbol e}_j$, $\underline{\boldsymbol p}_k$, and $\underline{\boldsymbol c}_l$, respectively, in a tabular form. This is followed by generating an object \texttt{db} of class \texttt{database}, which connects to an empty \texttt{Neo4j} database and fills it with the tissue data according to the rules of the GVM's metagraph, using function \texttt{setup\_DB()}. The initialization is followed by a time loop, which propagates the system forward in time by time steps $\Delta t$. At each time step, the dynamical system [Eq.~(\ref{eq:dynSys})] is integrated between $t$ and $t+\Delta t$~[function \texttt{solver()}] and the program checks whether any of the edges in the cell network meets contitions for topological transitions~[function \texttt{topological\_transitions()}]. In particular, edge $j$ undergoes an ET transition if it is shorter than a threshold length $l_{\rm th}$ and the rate of change of its length is negative (${\rm d}l_j/{\rm d}t<0$), i.e., the edge is contracting. If the edge happens to be part of a triangle, a TE transition is performed on the triangle if, additionally, all edges of the triangle are shorter than $l_{\rm th}$ and the area of the triangle is decreasing (${\rm d}A_k/{\rm d}t<0$).

For every edge/triangle, subject to a topological transition, the core program sends a sequence of \texttt{Cypher} queries to the \texttt{Neo4j} graph database. These queries (i)~perform pattern matching to isolate a subgraph relevant for the particular transformation being performed, and (ii)~perform graph transformation on that subgraph. After graph transformations, vertex positions are displaced such that the lengths of the newly created edges (edges $e_7$, $e_8$, and $e_9$ in Fig~1) are on the order of $l_{\rm new}\ll 1$. Finally, the local structure of the arrays, encoding $\underline{\boldsymbol r}_i$, $\underline{\boldsymbol e}_j$, $\underline{\boldsymbol p}_k$, and $\underline{\boldsymbol c}_l$, (stored in object \texttt{t}) are updated according to the applied transformations. This is done by converting the altered part of the knowledge graph back into the array format using function \texttt{update()}.

\section*{Acknowledgments}
We thank Urban \v Zeleznik for generating the disordered initial configurations for benchmarking. We thank Tomer Stern, Domen Vaupoti\v c, Sta\v s Adam, and all members of the Theoretical Biophysics Group at Jožef Stefan Institute for fruitful discussions. The main idea for this project surfaced from a very fruitful collaboration with Wisdom Labs. Thanks to all Wisdom Labs folks, especially Andra\v z Tu\v s, Kristjan Pe\v canac, Luka Stopar, Marko Zadravec, Jasna Pe\v canac, and Jan Lampi\v c.















%
%
%

\newpage
\section*{Supporting information}
%
\setcounter{figure}{0}
\renewcommand{\thefigure}{S\arabic{figure}}

\paragraph*{S1 Fig.}
\label{S1_Fig}
{\bf Graph transformations of EV and VT transitions.}~(panels  {\bf A} and {\bf B}, respectively). Graphs in the left and the right column correspond to the initial and the final cell configuration, respectively. Gray arrows represent relationships labeled \texttt{IS\_PART\_OF}. The graphs in the middle column show graph transformations, which include green and red relationships, indicating relationship creations and deletions, respectively. Additionally, the graph transformation specifies property values of the newly created relationships. In each graph, the node indices increase from left to right in unit steps.

\paragraph*{S2 Fig.}
\label{S2_Fig}
{\bf Graph transformations of TV and VE transitions.}~(panels {\bf A} and {\bf B}, respectively). Graphs in the left and the right column correspond to the initial and the final cell configuration, respectively. Gray arrows represent relationships labeled \texttt{IS\_PART\_OF}. The graphs in the middle column show graph transformations, which include green and red relationships, indicating relationship creations and deletions, respectively. Additionally, the graph transformation specifies property values of the newly created relationships. In each graph, the node indices increase from left to right in unit steps.

\paragraph*{S3 Fig.}
\label{S3_Fig}
{\bf The algorithm underlying \texttt{neoVM}.}~A 3D cell aggregate is set up from an input \texttt{.vt3d} file by function \texttt{setup\_from\_vt3d()} and then converted into a graph database, set up in \texttt{Neo4j} by function \texttt{setupDB()}. These  initialization steps then followed by a time loop, which iterates between \texttt{solver()} and \texttt{topological\_transitions()} functions. The function \texttt{solver()} calculates geometric properties of cells and the associated gradients (i.e., conservative forces) and propagates the system forward in time. The function \texttt{topological\_transitions()} loops over all cell edges to find those that meet criteria for topological transitions. For edges that meet these criteria, topological transitions are performed through pattern matching and graph transformations (functions \texttt{pattern\_matching()} and \texttt{graph\_transformation()}, respectively), applied directly to the graph database. Finally, the tissue is updated accordingly by function \texttt{update()} and positions of vertices involved in the topological transformation are corrected by function \texttt{correct\_vertex\_positions()}.

\paragraph*{S4 Fig.}
\label{S4_Fig}
{\bf Performance of \texttt{neoVM}.}~Runtime duration per unit simulation time as a function of cell count. Here, two separate data sets represent two strengths of active fluctuations, i.e., $\sigma = 0.2$ and $\sigma = 0.4$, shown in red and green dots, respectively. In both cases, simulation runtime shows a linear dependence (in a blue dashed lines) on the size of cell aggregates.

\end{document}